# Volador 1.0: A Data-Driven Air-Sea Full-Coupling Regional Forecast Model with Submesoscale-Permitting Based on MOE-Swin-Transformer Framework

Yuhang Zhu[1+], Jianxin Wang[2+], Yu-kun Qian[1], Yineng Li[1], Yahui Liu[2], Yankun Gong[1], Shilin Tang[1], Shiqiu Peng[1*], Tao Song[2*]

[1]Laboratory of Tropical Oceanography, South China Sea Institute of Oceanology, Chinese Academy of Sciences, Guangzhou, 510301, China.

[2]Qingdao Institute of Software, College of Computer Science and Technology, China University of Petroleum (East China), Qingdao, 266580, China

* Corresponding authors: Shiqiu Peng (speng@scsio.ac.cn) and Tao Song (tsong@upc.edu.cn)

## Abstract

A data-driven air-sea full-coupling regional forecast model with submesoscale-permitting, named “Volador 1.0”, is developed for the South China Sea (SCS). The model features a Swin-Transformer framework integrated with a Mixture-of-Experts (MoE) system, a latent space interaction architecture based on Cross-Grid Bidirectional Cross-Attention, and a fast-slow dual-branch architecture. Both the three-month hindcast test and the 15-day operational real-time forecasting demonstrate that Volador 1.0 has a very encouraging and promising performance in 0-72h forecasting of temperature and salinity in the 0-500m upper ocean as well as the sea surface height with root-mean-square-error (RMSE) or mean absolute error (MAE) smaller than or at least comparable to those from the reanalysis datasets REDOS V2.0 and GLORYS12 and the state-of-the-art regional numerical model Regional Ocean Modeling System (ROMS). In particular, Volador 1.0 demonstrates its capability of capturing/forecasting submesoscale processes including internal waves, with an energy spectrum well representing sub- to mesoscale energy cascade as expected by the classical turbulence theory. Further analysis based on ablation experiments shows that the air-sea full-coupling framework, which takes into account the dynamic exchanges of momentum and heat fluxes between the atmosphere and the ocean, indeed helps improve the model’s performance compared to the non-full-coupling one. Volador 1.0, though still subject to refinement in the coming future with a large space for improvement, blazes a path for an accurate, fine and fast marine environment forecasting, and thus could help promote our capability of disaster prevention and mitigation in the SCS as well as in other coastal regions where these innovative techniques can be applied.

**Keywords:** Air-sea full-coupling; Submesoscale-permitting; Swin-Transformer; Mixture-of-Experts (MoE); Marine forecasting; Data-driven.

## 1 Introduction

The South China Sea (SCS), which connects the Pacific and Indian Oceans, is a semi-enclosed deep-water basin with an area of 3.5 million square kilometers. Its bathymetry is characterized by a deep central basin, bordered to the north and south by two broad shallow shelves and to the west and east by two steep continental slopes (Qu 2000; Wang et al. 2006). Dominated by the East Asian monsoon, its circulation reverses seasonally between a cyclonic pattern in winter and an anti-cyclonic one in summer (Fang et al. 1998; Hu et al. 2000; Liu et al. 2008). Further complexity arises from dynamical and physical processes, such as the western boundary current (Wang et al. 2013), the SCS throughflow (Wang et al. 2006; Song 2006; Yu et al. 2007), the mesoscale eddies (Li et al. 2011; Chen et al. 2011, 2012), and the active internal tides and waves (Guo et al. 2012; Ma et al. 2013; Zhao. 2014), and so on. The complex interplay among its geographical setting, monsoon climate and dynamical processes makes marine forecasting in the SCS a significant challenge.

Numerical models coupled with data assimilation systems have long been the most commonly used and well-established method for marine forecasting (e.g., Smedstad et al. 2003; Chassignet et al. 2007; Miyazawa et al. 2017). Numerical models simulate spatial and temporal variations of the ocean by solving the partial differential equations which approximately describe the ocean state. With the support of data assimilation that incorporates real oceanic observations into the initial field by solving the cost function, the marine simulation and forecasting based on numerical modelling have achieved a big success. However, numerical modeling and data assimilation typically demand substantial computational resources and tend to be slow in calculation. And as oceanographic research and forecasting demand shift towards higher spatiotemporal resolution (e.g., diurnal and submesoscale processes), the computational cost and efficiency of numerical models and data assimilation systems face even greater challenges in practical application.

In recent years, the rapid advancement of Artificial Intelligence (AI) has offered a new paradigm in the oceanic research. An increasing number of AI-based methods are widely applied to marine simulation and forecasting (e.g., Song et al. 2021; Lam et al. 2023; Bi et al. 2023). Compared to numerical models, AI-based simulation and forecast models can automatically learn spatiotemporal relationship from vast amounts of oceanic datasets, effectively capturing the pattern of oceanic variability without requiring prior knowledge of physical mechanism. They not only rival conventional numerical models in forecast accuracy but also demand significantly lower computational resources (Bi et al. 2023). Currently, an increasing number of AI-based big models for marine forecasting are being developed (e.g. Xiong et al. 2023; Wang et al. 2024; Aouni et al. 2024; Cui et al. 2025) with promising results in forecast accuracy, highlighting the revolutionary potential of AI for marine forecasting.

However, the development of AI-based ocean models is still in its infancy with substantial room for improvement. For instance, the training data for most of the current AI-based big ocean models are primarily sourced from the GLORYS12 reanalysis from the Copernicus Marine Environment Monitoring Service (CMEMS, Jean-Michel et al. 2021), or the HYCOM reanalysis dataset from the U.S. Naval Research Laboratory (Cummings et al., 2013), both of which are available with the horizontal scale of 1/12°. Limited by the spatiotemporal resolution of the training data, the resolving scales of these models are up to those of mesoscale eddies, let alone submesoscale features unresolvable. Beyond this, none of them realizes a two-way coupling between the atmospheric and oceanic components in the overall model training and inference, i.e., the atmospheric state is fixed all the time, which inevitably induces biases due to the neglection of the dynamic exchanges of momentum and heat fluxes between the two components.

To address these issues, we developed a data-driven air-sea full-coupling regional forecast model with submesoscale-permitting for the SCS and the Northwestern Pacific Ocean (NWPO, 99°E~134°E, 1°N~30°N, Figure 1), named "Volador 1.0". A volador, with a light and nimble body, can not only swim in the sea but also fly in the air very fast, which perfectly matches the unique features of our model, i.e., air-sea full-coupling, light-weight, and high speed. And in the tales of ancient China, a volador could dispel some strange illnesses of human beings and thus bring people good fortune, which also highly aligns with the utility of the model we pursue and expect that can help people prevent hazards and reduce property or life losses in coastal regions.

The structure of this paper is as follows: Section 2 introduces the data used for training and inference of Volador 1.0. Section 3 describes the architecture of Volador 1.0. The evaluation of Volador 1.0’s performance is presented in Section 4. Section 5 gives a summary.

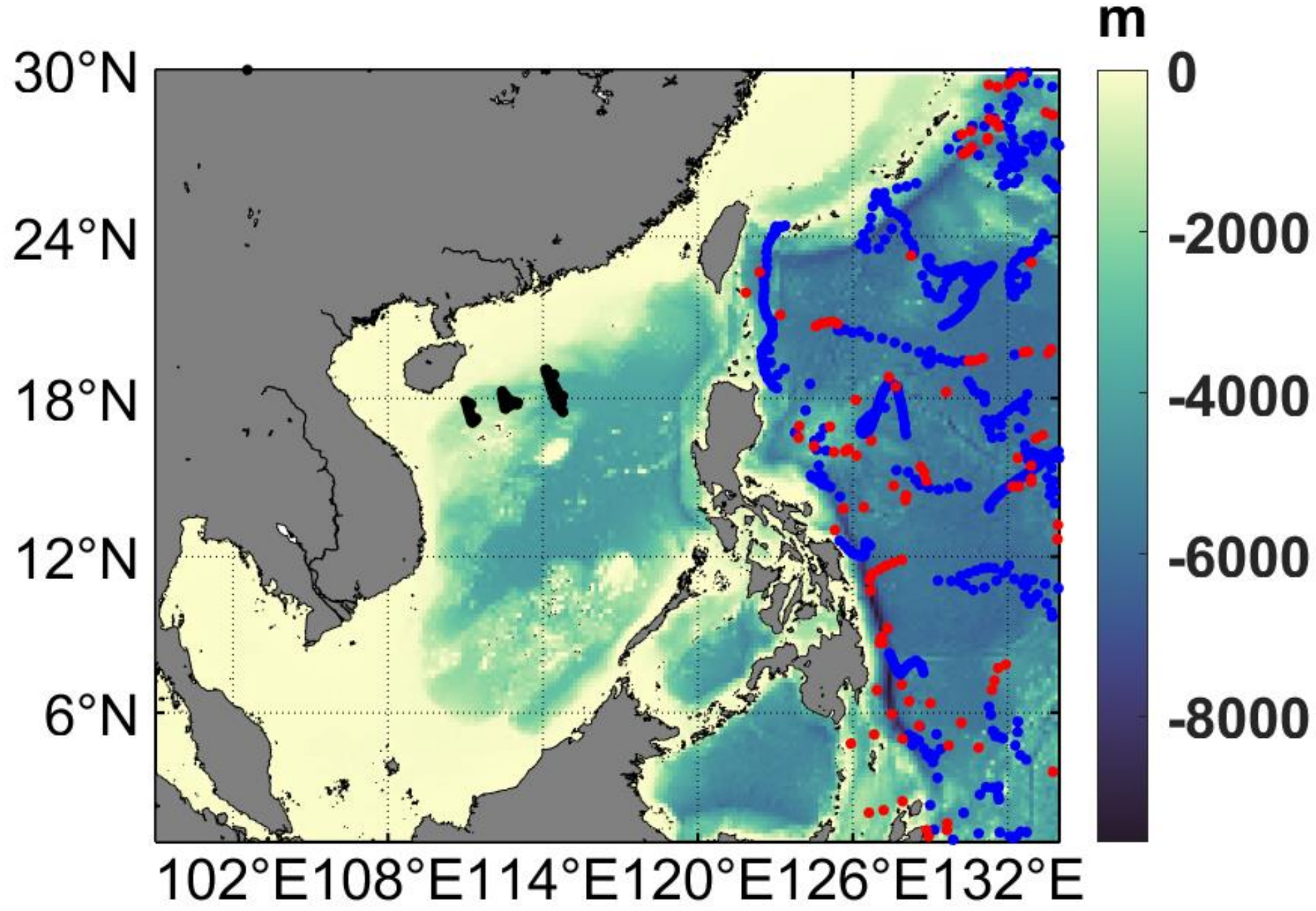

Figure 1 Model domain and topography for Volador 1.0. Black and blue dots indicate the locations of underwater gliders and Argo buoys used to evaluate T/S in the three-month hindcast test, and red dots indicate the locations of Argo buoys used to evaluate T/S in the operational real-time forecasting.

## 2 Data

The ERA5 dataset, which is widely regarded as the most comprehensive and accurate reanalysis, is used for training/inferring the atmospheric module of Volador 1.0. The subset of the ERA5 dataset with the time span of Jan. 1, 2020-Aug. 31, 2024, the domain of 99°E~134°E, 1°N~30°N, the spatiotemporal resolution of 1/4°×1/4° and 6 hours is adopted. In Volador 1.0, the atmospheric variables, including the temperature at 2m (*t2m*), mean sea level pressures (*msl*), u-wind and v-wind at 10m (*u10* and *v10*), specific humidity (*q*) and geopotential (*z*), are utilize for training/inference.

The Reanalysis Dataset of the South China Sea (REDOS) V2.0, developed by the South China Sea Institute of Oceanology, is used for training/inferring the oceanic module of Volador 1.0. REDOS V2.0 was developed based on the Regional Ocean Modeling System (ROMS) and a Multi-Scale Three-Dimensional Variational (MS-3DVAR) assimilation system. The temporal coverage spans from January 1, 1992, to December 31, 2024, totaling 33 years with temporal resolution of 1 hour. The spatial domain extends from 1°N to 30°N and 99°E to 134°E with horizontal resolution of 1/36°×1/36° and vertical layers of 40 (0-5000m), encompassing the entire SCS and parts of the NWPO. During the generation of REDOS V2.0, atmospheric forcing was provided by the ERA5 with time interval of 6 hours, and lateral boundary conditions were derived from the GLORYS12 (Lellouche. 2021) with temporal interval of 5 days. The tidal forcings on the lateral boundary are the tidal level and currents calculated through 10 main tidal components (M2、S2、N2、K2、K1、O1、P1、Q1、M4、MS4) provided by the TPXO8-atlas with a horizontal resolution of 1/30° ×1/30° (Egbert and Erofeeva, 2002). Besides, the monthly river discharge forcings from 10 main runoffs around the SCS (including the Mekong, Xijiang, Rajang, Beijiang, Pahang, Chao Phraya, Dongjiang, Kinabatangan, Kelantan and Oujiang) provided by Dai et al (2017) are implementated. The observations assimilated in the REDOS V2.0 include the gridded satellite-derived sea level anomalies (SLA), gridded satellite-derived sea surface temperature (SST), as well as the temperature and salinity (T/S) profiles from Argo buoys (Zhang et al. 2018), the World Ocean Database 2023 (WOD23, Mishonov et al. 2023), the CTD, XBT and underwater gliders from research cruises conducted by the SCSIO. Additionally, gridded T/S profiles inverted by the MODAS (0-2000m) (Mao et al., 2013) and extracted from the World Ocean Atlas 2023 (WOA 2023, beneath 2000m) (Reagan et al. 2024) are also assimilated to further correcting the large scale T/S biases. In Volador 1.0, the subset of the REDOS V2.0 with the same time span, temporal resolution and domain as the ERA5 is adopted, including the variables of temperature (*temp*), salinity (*salt*), u-current (*u*), v-current (*v*) from sea surface to the depth of 500m, and sea surface height (*SSH*).

For the model evaluation of Volador 1.0, the T/S profiles measured by SCSIO-deployed underwater gliders and by Argo buoys provided through China Argo Data Center, and the SWOT (Surface Water and Ocean Topography) along-track sea level anomaly (SLA) data provided through Physical Oceanography Distributed Data Active Archive Center, National Aeronautics and Space Administration ((PO.DAAC, NASA), are used.

# 3 Methodology

## 3.1 Overall Architecture of Volador 1.0

The overall architecture of Volador 1.0 can be formulated as a unified, end-to-end deep learning framework capable of fitting a high-dimensional mapping function:

$$F: (X_{ocean}, X_{atmos}, \tau, \lambda) \rightarrow Y_{pred}$$

where $X$ represents the atmospheric and oceanic inputs, $\tau$ and $\lambda$ represent the encoding parameter and forecast lead time parameter, respectively, and $Y_{pred}$ is the forecasting outputs.

As illustrated in Figure 2, Volador 1.0 first applies channel-wise Z-score normalization to the input four-time-step oceanic and atmospheric data to prevent the varying dimensions of different forecasting variables from affecting the data distribution. Then, it feeds the processed data into a Patch module, which uses convolution operations to halve the spatial length and width dimensions of the input data, effectively extracting structural data features and reducing the overall dimensionality to 548×632×96. Next, the data enters a three-layer Swin-Transformer backbone network integrated with a Mixture-of-Experts (MoE) system, which utilizes downsampling and upsampling modules as intermediate connections to help the model learn large-scale features and restore the original data scales of the oceanic and atmospheric variables. In contrast to conventional methods relying on bulk formulas to directly calculate air-sea fluxes, Volador 1.0 adopts a high-dimensional implicit coupling mechanism within its architecture, which enables the model to autonomously learn non-linear momentum and heat flux interaction processes directly within the feature space. Furthermore, Volador 1.0 introduces a sinusoidal absolute time encoding parameter $\tau$ and a forecasting lead time parameter $\lambda$, which can effectively capture the seasonal and diurnal cyclic variations of the training data, thus achieving multi-step end-to-end forecasting with avoidance of the inherent error accumulation problem found in iterative autoregression. Finally, Volador 1.0 generates a standardized residual sequence $\Delta O$, corresponding to an output $Y_{pred}$ at forecasting lead time during a single forward pass, and then denormalizes the output residual variables and superimposes them onto the initial state, reconstructing the complete atmospheric and oceanic forecasting fields. A detailed introduction of the modules of Volador 1.0 is as follows.

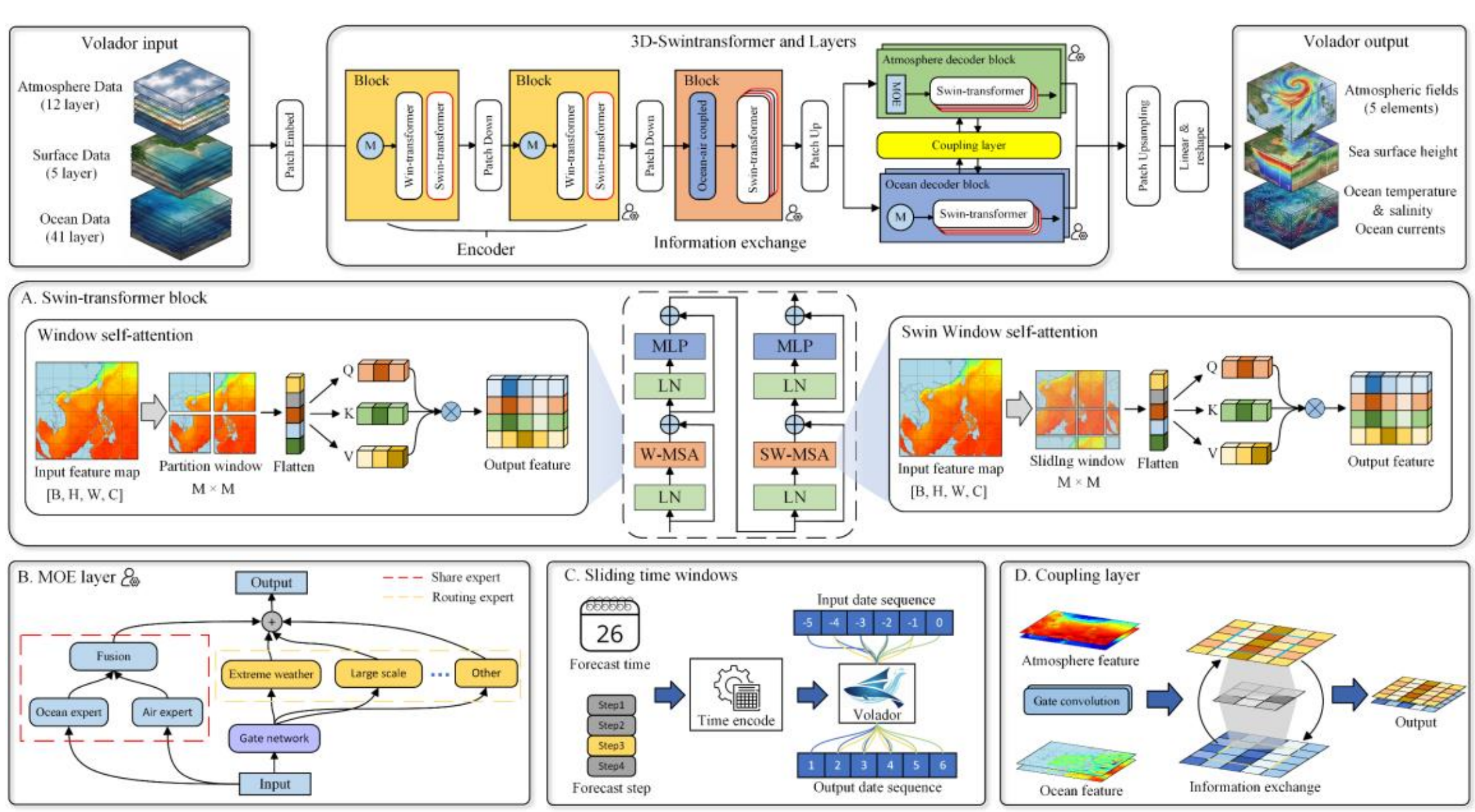

Figure 2 The overall architecture of the Volador 1.0. The top panel is the end-to-end pipeline, processing multi-source atmospheric and oceanic inputs through a 3D-Swin-Transformer encoder and reconstructing them via a parallel dual-branch decoder. (A) Swin-Transformer block: the core feature learning backbone, which replaces the conventional feed-forward network with an adaptive Mixture-of-Experts (MoE) module. (B) MoE layer: a dynamic routing mechanism that activates specific expert sub-models to decouple complex ocean dynamic tasks. (C) Sliding time windows: incorporates historical sequences and time encoding to enable continuous multi-step forecasting. (D) Coupling Layer: a purely data-driven latent space interaction module based on Cross-Grid Bidirectional Cross-Attention, utilizing dynamic grid sampling to achieve the alignment and deep bidirectional fusion of air-sea features with heterogeneous resolutions.

### a. Encoder and Decoder Processing

The encoder consists of three Swin-Transformer blocks in the first half of Volador 1.0. As the network depth increases, the encoder scales the feature information from submesoscale details to large-scale circulations. The feed-forward neural network in the final encoder block is replaced with an adaptive MoE system to enhance the model's encoding capabilities across different environments. The Swin-Transformer module, which forms the fundamental backbone for the model's feature learning, is designed as a multi-head self-attention mechanism that combines regular and shifted windows, and effectively captures the multi-scale spatial correlations of ocean dynamic processes and solves the computational resource problems caused by high-resolution data. Notably, different oceanic variables exhibit significantly distinct physical evolution rates. Specifically, the sea temperature and salinity, functioning as thermodynamic scalars, are characterized by relatively slow evolution and smooth spatial distribution; in contrast, tide-included current and SSH fields contain more transient high-frequency variations and sharp dynamic structures. The commonly used single decoding mechanism easily causes gradient conflicts and feature over-smoothing problems when reconstructing these disparate variables simultaneously. Therefore, we innovatively introduce a fast-slow dual-branch architecture based on physical quantity decoupling within the decoder stage by splitting the feature reconstruction process into two parallel computational pathways:

$$\hat{Y}_{TS} = D_{slow}\left(Z_{deep};\ \theta_{slow}\right),\ \hat{Y}_{UV} = D_{fast}\left(Z_{deep};\ \theta_{fast}\right)$$

where $\hat{Y}_{TS}$ and $\hat{Y}_{UV}$ denote the predicted outputs for the slow-varying scalar variables and the fast-varying dynamic vector variables, respectively. $Z_{deep}$ represents the deep latent features extracted from the encoder bottleneck. $D_{slow}$ and $D_{fast}$ represent the scalar and vector branches parameterized by $Z_{deep}$, $\theta_{slow}$ and $\theta_{fast}$. $D_{slow}$ focuses on reconstructing the continuity and structural features of slow-varying elements, such as temperature, salinity and SSH fields, while $D_{fast}$ specifically optimizes the high-frequency dynamic evolution and fine edges of the current. This physically inspired dual-branch decoupling design allows Volador 1.0 model to adaptively match the spatiotemporal evolution patterns of different oceanic elements during the feature mapping process, effectively eliminates mutual interference during multi-task learning, and thus significantly improves the reconstruction accuracy of complex ocean dynamic features. Furthermore, we embed an air-sea full-coupling module within the encoder's bottleneck region where encoded features pass to the downstream decoding module, thereby establishing cross-modal feature interactions between the ocean and the atmosphere within the latent space.

**b. Air-Sea Full-Coupling Module**

For depicting the air-sea momentum and heat interactions which are crucial for oceanic and atmospheric evolution, a novel cross-modal air-sea full-coupling module is implemented within the deep bottleneck stage where encoder features propagate downstream. This module innovatively proposes a latent space interaction architecture based on Cross-Grid Bidirectional Cross-Attention, establishing an efficient information exchange channel for dynamically forecasting atmospheric and oceanic states. This purely data-driven implicit coupling paradigm not only avoids the error accumulation inherent in conventional parameterization schemes, but also provides the ocean system with dynamic constraints that strictly adhere to real fluid boundary conditions.

The design details of the air-sea full-coupling module are as follows:

First, to address the spatio-temporal heterogeneity in grid resolutions between ocean and atmosphere, the air-sea full-coupling module introduces a cross-grid sparse partitioning strategy. This design constrains the quadratic complexity of global attention to local grid spaces. Furthermore, without relying on any spatial interpolation, it resolves the length mismatch between oceanic and atmospheric sequences, achieving seamless alignment of heterogeneous grids along the feature sequence dimension.

Subsequently, within a uniformly mapped embedding dimension, the air-sea full-coupling module utilizes oceanic and atmospheric features mutually as Queries and Key-Values to perform parallel bidirectional feature learning in the latent space.

Finally, the air-sea full-coupling module spatially reconstructs the interacted feature sequences and superimposes them onto the original feature maps via residual connections. To ensure physical and dynamical consistency, strict geographical mask constraints are applied at the output of the coupled features. This mask mathematically enforces the isolation of land-sea boundaries during computation, ensuring that atmospheric signals within land grids do not spill over into oceanic regions.

### c. Adaptive MoE System for Oceanic Processes

Differring from the current AI models that use single network weights to represent air-sea multiscale physical processes, Volador 1.0 innovatively introduces an adaptive MoE architecture during the deep feature encoding and decoding stages. Specifically, Volador 1.0 replaces the feed-forward network layer in the conventional Swin-Transformer module with a multi-expert module cluster led by a gating network, thereby decoupling the complex ocean dynamic system forecasting task into different expert sub-tasks.

During feature forward propagation, the dynamic routing evaluates the local dynamic state of the current input feature in real-time, and calculates sparse weights and selects the optimal expert models to process the features based on the weight ranking. This physically inspired task decoupling strategy enables different expert modules to directionally learn specialized perception and thus has calculation capabilities for specific sea conditions during training. Besides, the MoE architecture leverages its sparse activation characteristics to require only a minimal number of activation parameters during the inference stage, thereby achieving fine-grained forecasting of multi-scale, highly non-linear ocean dynamical processes and significantly reducing inference power consumption.

**d. Dynamic Land-Sea Mask**

Volador 1.0 model constructs a multi-layer dynamic land-sea mask mechanism to ensure that the extracted three-dimensional dynamic features strictly comply with the actual ocean depth distribution. Independent mask constraints that match the real topography are implemented for each of the ten vertical depth layers of the oceanic model, respectively, thereby precisely restricting the learning of water body dynamic features at different depths within the valid ocean grids. Furthermore, the forced truncation of the land mask often generates computational artifacts in the model's numerical gradients at the land-sea boundary zones, which thus affects the receptive field of the convolution process during deep learning. Therefore, we innovatively adopt a land-sea soft mask strategy in the model's attention aggregation. By dynamically adjusting the bias term of the self-attention matrix, this strategy effectively reduces the feature interference from land grids on ocean grids in the boundary areas. The soft mask design enables the model to maintain the smoothness of feature learning at land-sea boundaries, and successfully avoids high-frequency oscillations caused by hard truncation, thereby forcing the neural network to focus its representation learning attention weights on the valid oceanic regions.

**e. Training and Fine-tuning**

In order to balance the generalized expression of physical atmospheric and oceanic laws with the sensitive capture of extreme high-impact atmospheric and oceanic events, Volador 1.0 adopts a two-stage training strategy, consisting of pre-training and fine-tuning, for the model's parameter optimization. During the pre-training stage, an end-to-end training process is conducted based on the atmospheric dataset ERA5 and oceanic dataset REDOS V2.0 with 6-hour intervals as introduced in section 2. This stage aims to help the network implicitly deduce the short-term fluid dynamic evolution patterns of the air-sea full-coupling system from massive long-term time-series data. During the pre-training state, the global parameter $\theta$ is optimized by minimizing the forecasting residuals across global spatiotemporal samples:

$$L_{pre}(\theta) = E_{(X,Y)\sim D_{global}}[\|F(X;\theta) - Y\|^2]$$

where $D_{global}$ represents the atmospheric and oceanic training datasets, and $F(\cdot)$ denotes the model's forward mapping function. Due to the long-tail distribution of extreme high-impact atmospheric and oceanic events (such as passing typhoons) within the complete datasets, conventional global optimization often smooths out the steep numerical gradients generated by extreme conditions. Therefore, we specifically filter and construct a sliced dataset $D_{ext}$ that contains intense air-sea thermodynamic exchange processes (such as typhoons) for the fine-tuning stage. Furthermore, to avoid the degradation of basic air-sea dynamic features due to catastrophic forgetting, the expert module is optimized from pre-trained parameters with a lower learning rate $\beta$ during the fine-tuning stage, thereby strictly constraining the update step size of the network weights:

$$\theta_{t+1} = \theta_t - \beta\nabla_\theta L_{ext}(F(X_{ext};\theta_t), Y_{ext})$$

where $\theta_t$ and $\theta_{t+1}$ denote the network parameters at training iteration $t$ and $t+1$, respectively. $X_{ext}$ and $Y_{ext}$ represent the input features and the corresponding ground truth targets sampled from the extreme event dataset $D_{ext}$. $F(X_{ext};\theta_t)$ indicates the model's prediction given the parameters at step $t$, $L_{ext}$ signifies the loss function evaluated specifically on this sliced dataset, and $\nabla_\theta$ represents the gradient operator. This two-stage training strategy smoothly transitions from generalization to specialization, ensuring the model's physical consistency during forecasting and help improving its forecasting accuracy under extreme conditions.

### 3.2 Input data for training and testing Volador 1.0

The atmospheric data, which includes variables of *t2m*, *msl*, *u10*, *v10*, *q* and *z* near the sea surface, and the oceanic data that contains *temp*, *salt*, *u*, *v* from sea surface to depth of 500m (totally 10 layer) and *SSH* at the time step of T0, T0-6h, T0-12h and T0-18h (where T0 denotes the initial time of forecasting, and '-Xh' means 'X' hours preceding T0) are used as input of Volador 1.0, forming two data tensors with sizes of $N_{lat}^a \times N_{lon}^a \times N_t$ and $N_{lat}^o \times N_{lon}^o \times N_{lev} \times N_t$, respectively, where $N_{lat}^a = 121$, $N_{lon}^a = 141$, $N_{lat}^o = 1096$, $N_{lon}^o = 1264$, $N_{lev} = 10$ and $N_t = 4$.

Currently, due to limited computation resource, only 4.5-year data from Jan.1, 2020 to May 31, 2024 are used for training Volador 1.0, and the three-month data from Jun. 1, 2024 to Aug. 31, 2024 are used for testing the model. An amount of approximately 130 million learnable parameters is involved in training and testing.

# 4 Evaluation on the performance of Volador 1.0

## 4.1 Performance of Volador 1.0 in a three-month hindcast test

In this study, we focus our evaluation primarily on the performance of Volador 1.0 in forecasting the ocean state, especially in capturing and forecasting the submesoscale processes of the ocean. For this purpose, a three-month hindcast test is conducted from June 1, 2024 to August 31, 2024. The initial fields of atmosphere and ocean are provided by the EAR5 and REDOS V2.0 dataset, as introduced in section 2 and 3.2. The experimental setup employs a daily rolling forecast scheme, initialized at 0000 UTC of each day, which outputs atmospheric and oceanic states for the upcoming 72 hours with 6-h interval or 1-h interval (only for SSH).

### a. Overall performance in marine forecasting

The overall performance of Volador 1.0 in forecasting the ocean state is evaluated by comparing the forecasted T/S profiles against those measured by Argo buoys in the NWPO and by SCSIO-deployed underwater gliders in the northern SCS (Figure 1), with additional comparison to those from REDOS V2.0 and GLORYS12 reanalysis datasets. The daily averaged T/S profiles fromVolador 1.0 and the reanalysis datasets are interpolated onto the locations of observed ones according to the corresponding observation date, and the root-mean-square-errors (RMSEs) of T/S are then calculated at each vertical layer:

$$\text{RMSE} = \sqrt{\frac{1}{n}\sum_{i=1}^{n}(fcst_i - obs_i)^2}$$

where $fcst_i$ and $obs_i$ represent the forecasts and observations at $i$th point, respectively, and $n$ is the number of observed T/S profile samples. Figures 3 gives the RMSEs of T/S profiles for 0-24h, 25-48h and 49-72h forecast periods from Volador 1.0 forecasts and from REDOS V2.0 and GLORYS12 datasets, and the corresponding vertically-averaged RMSEs are shown in Table 1. It can be seen that, probably benefiting from the higher T/S accuracy of REDOS V2.0 as input data, the RMSEs of T/S profiles fromVolador 1.0 in both the SCS and the NWPO are all smaller than those from GLORYS12 at all forecast periods, and are also smaller than those from REDOS V2.0 at forecast period of 0-24h. As expected, the RMSEs of T/S profiles increase with the forecasting period, but the increase is very slow, i.e., only from 0.521℃ (0.102PSU) at 0-24h forecast period to 0.614℃ (0.110PSU) at 49-72h forecast period, demonstrating the stability of Volador 1.0 in the sea temperature and salinity forecasting.

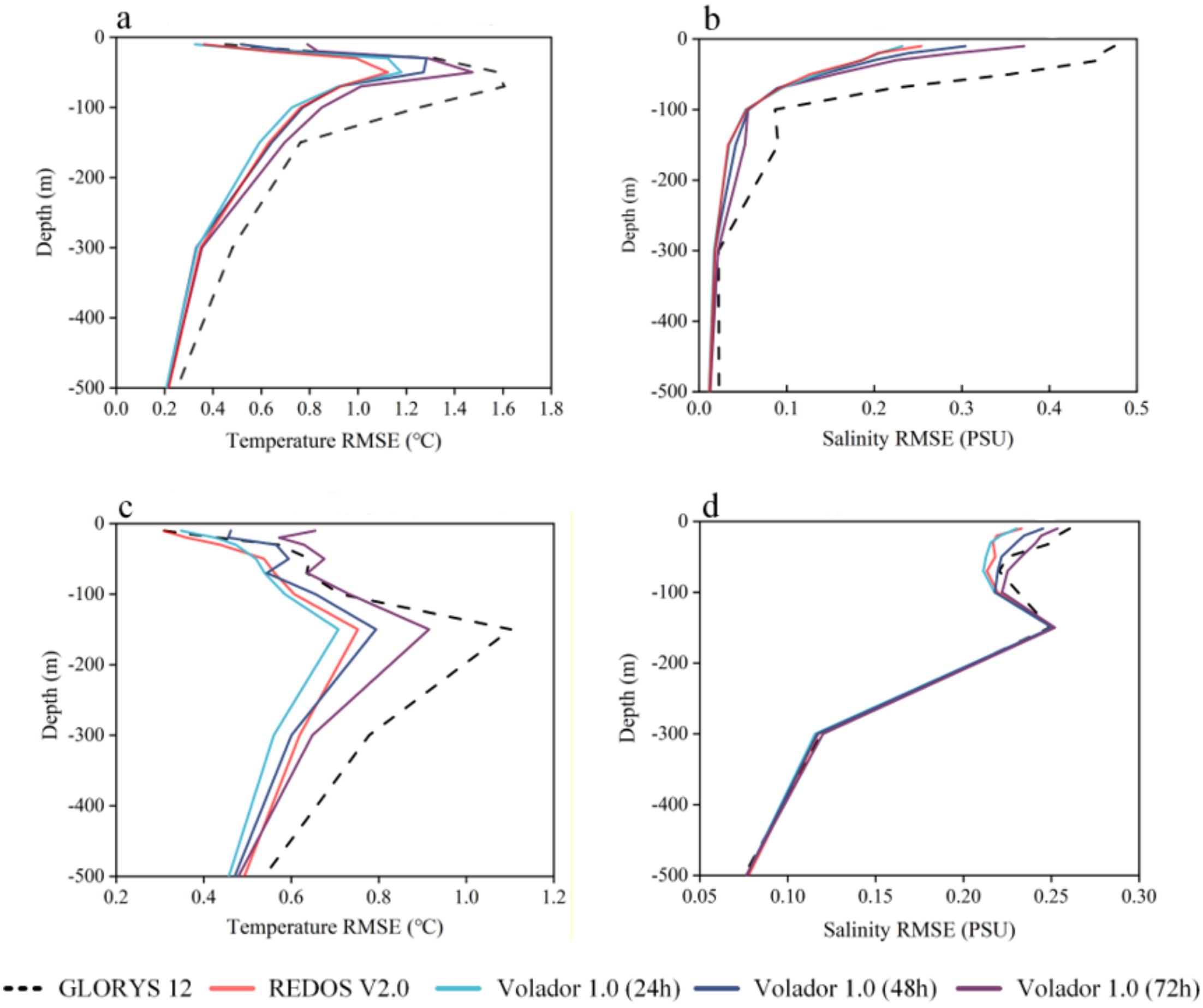

Figure 3 The domain-mean RMSEs of temperature (a, c) and salinity (b, d) profiles in the SCS (a, b) and the NWPO (c, d) from GLORYS reanalysis, REDOS V2.0 reanalysis, and Volador 1.0 with 0-24h, 25-48h and 49-72h forecasts averaged from June 1, 2024 to August 31, 2024.

Table 1 The domain-mean vertical-averaged RMSEs of temperature and salinity from GLORYS reanalysis, REDOS V2.0 reanalysis, and Volador 1.0 with 0-24h, 25-48h and 49-72h forecasts averaged from June 1, 2024 to August 31, 2024.

| Variable and Region \ Dataset | | REDOS V2.0 | GLORYS12 | VOLADOR 1.0 24h | VOLADOR 1.0 48h | VOLADOR 1.0 72h |
|---|---|---|---|---|---|---|
| Temperature (℃) | South China Sea | 0.496 | 0.682 | 0.484 | 0.510 | 0.557 |
| | Western Pacific | 0.595 | 0.752 | 0.558 | 0.606 | 0.671 |
| | All domain | 0.546 | 0.717 | 0.521 | 0.558 | 0.614 |
| Salinity (PSU) | South China Sea | 0.044 | 0.095 | 0.044 | 0.048 | 0.056 |
| | Western Pacific | 0.160 | 0.164 | 0.159 | 0.161 | 0.164 |
| | All domain | 0.102 | 0.130 | 0.102 | 0.105 | 0.110 |

The SSH forecasted by Volador 1.0 is evaluated against the SWOT along-track SLA, and is also compared with REDOS V2.0 and GLORYS12 datasets. The SSH from all datasets are first averaged on a daily basis to remove high-frequency tidal signals, and then the long-term mean SSH is subtracted to obtain the daily-averaged SLA for evaluation. Then, the daily-averaged SLA from Volador 1.0 and the reanalysis datasets are all interpolated onto the along-track observational locations according to the corresponding date, and the mean absolute errors (MAE) are calculated as:

$$\mathrm{MAE} = \frac{1}{n}\sum_{i=1}^{n}|fcst_i - obs_i|$$

where $fcst_i$ and $obs_i$ represent the forecasts and observations at $i$th point, respectively, and $n$ is the number of observed SLA samples. Figure 4 shows the horizontal distribution and the time series of domain-averaged MAE of SLA from Volador 1.0 forecasts for different forecast periods, REDOS V2.0 and GLORYS12. It can be seen that the patterns of spatial distribution and the temporal evolution of SLA from Volador 1.0 forecasts for different forecast periods are similar to those from the reanalysis datasets REDOS V2.0 and GLORYS12. Specifically, the domain-averaged MAEs from Volador 1.0 for 0-24h, 25-48h, and 49-72h forecast periods are 0.08m, 0.07m, and 0.07, respectively, comparable to those from the two reanalysis datasets which are all 0.08m.

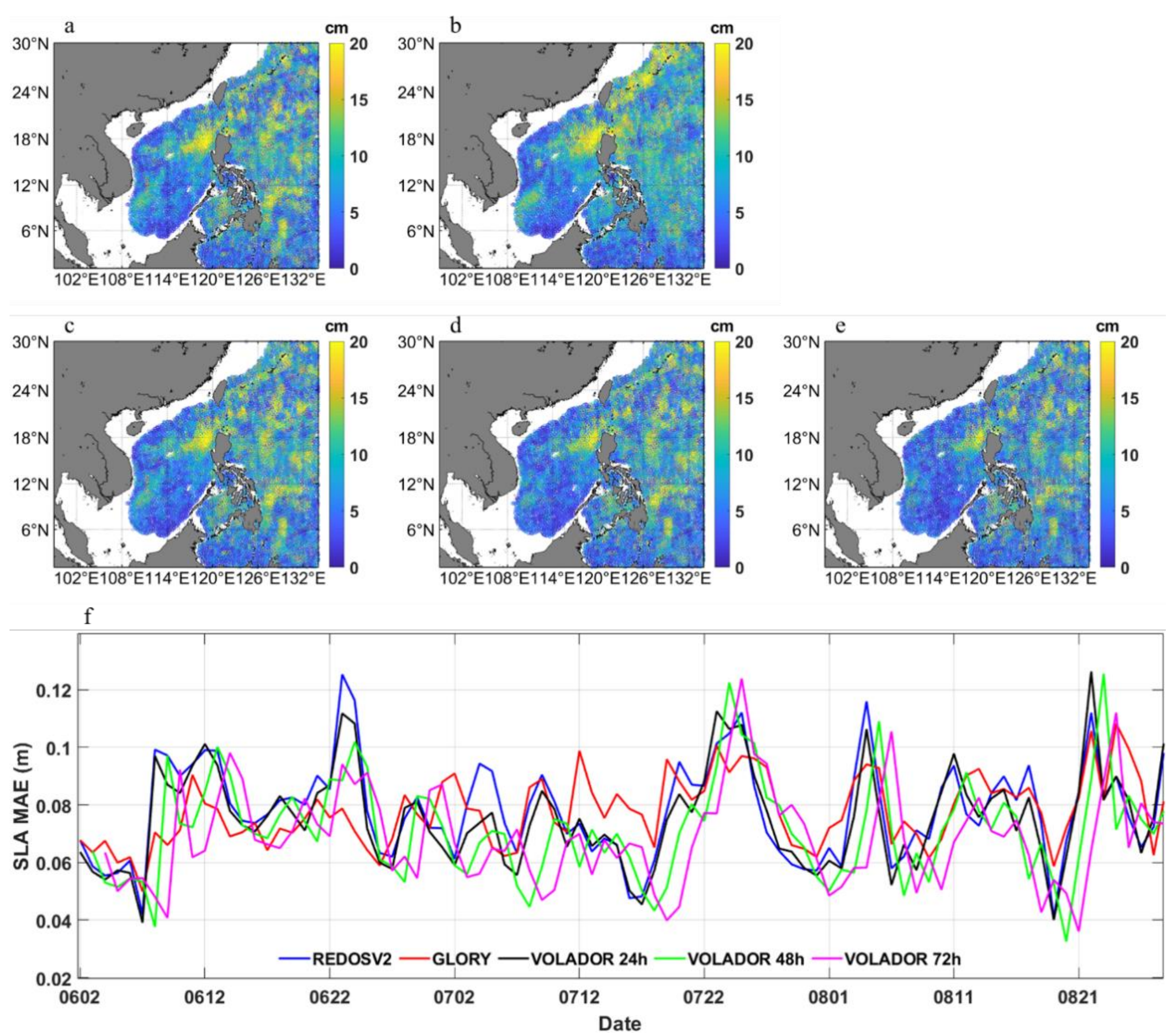

Figure 4 The horizontal distribution of time-mean MAEs (a, b ,c ,d ,e) and the time series (f) of domain-mean MAEs of SLA from REDOS V2.0 (a, f), GLORYS (b, f), and Volador 1.0 with 0-24h, 25-48h and 49-72h forecasts (c, d, e, f) averaged from June 1, 2024 to August 31, 2024.

**b. Capability of capturing/forecasting submesoscale features**

Submesoscale processes significantly influence air–sea interaction, mixed-layer dynamics, and the vertical transport of heat, salt, and nutrients (D'Asaro et al. 2011). Therefore, accurately representing ocean submesoscale information is crucial for a marine forecast system. Benefited from the high spatiotemporal resolution of the training dataset REDOS V2.0, Volador 1.0 possesses the potential capability of depicting and forecasting the submesoscale oceanic signals. In order to assess this capability, the Rossby number (Ro), which is a core diagnostic parameter for characterizing ocean submesoscale activity, and the internal waves, which share the same spatiotemporal scale of the submesoscale signals, are evaluated qualitatively against the independent REDOS V2.0 dataset.

Figure 5 shows the horizontal distribution of Ro from Volador 1.0 at the ending times of 24h, 48h and 72h forecasts initialized at 00:00 UTC on July 25, 2024 and that from REDOS V2.0 at the corresponding times. It can be seen that the Ro from Volador 1.0's forecasts at different forecast times exhibit spatial patterns similar to those from REDOS V2.0. There are numerous filament structures surrounding high- and low-vorticity centers associated with submesoscale eddies. In the shelf region of the northern SCS, a large number of elongated filament structures with high relative vorticity are present, which are generally distributed between eddies and some of them extend over a hundred kilometers. The results demonstrate that the Volador 1.0 is able to well capture submesoscale signals.

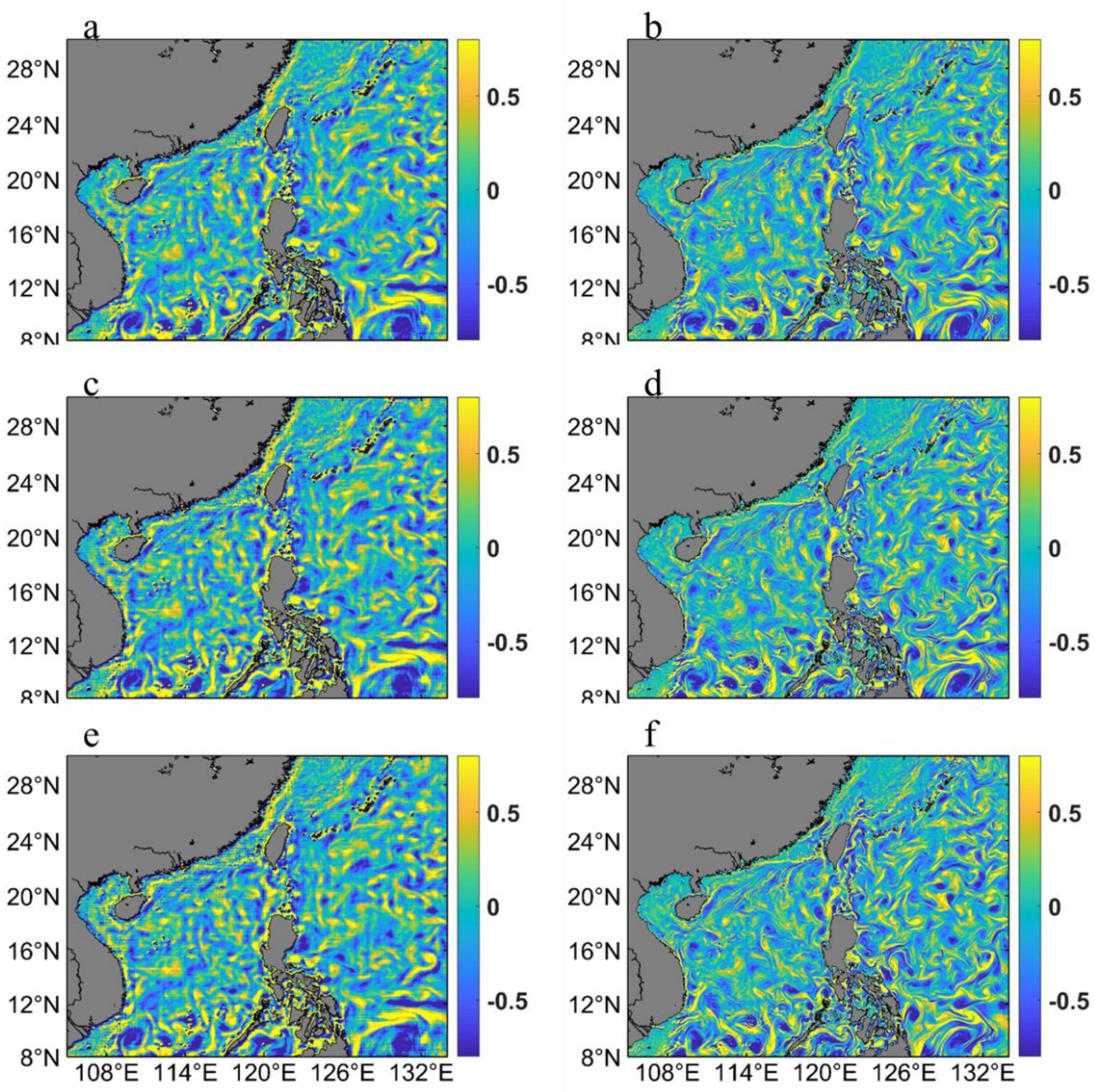

Figure 5 The horizontal distribution of Rossby number from Volador 1.0 (a, c, e) at the ending times of 24h, 48h and 72h forecasts (a, c, e) initialized at 00:00 UTC on July 25, 2024 of Volador 1.0, as well as those from REDOS V2.0 at the correponding

times (b, d, f).

The SSH gradients directly reflect the surface fluctuation pattern induced by internal waves. Figure 6 presents the SSH gradients calculated from Volador 1.0 at the ending times of 24h, 48h and 72h forecasts initialized at 00:00 UTC on July 25, 2024 and those from REDOS V2.0 at the corresponding times. It can be seen that, Volador 1.0 shows excellent performance in forecasting the internal waves regarding to the locations of crest lines when compared with those simulated by REDOS V2.0 even at the 72h forecast time, though the intensity of SSH gradients is relatively weaker.

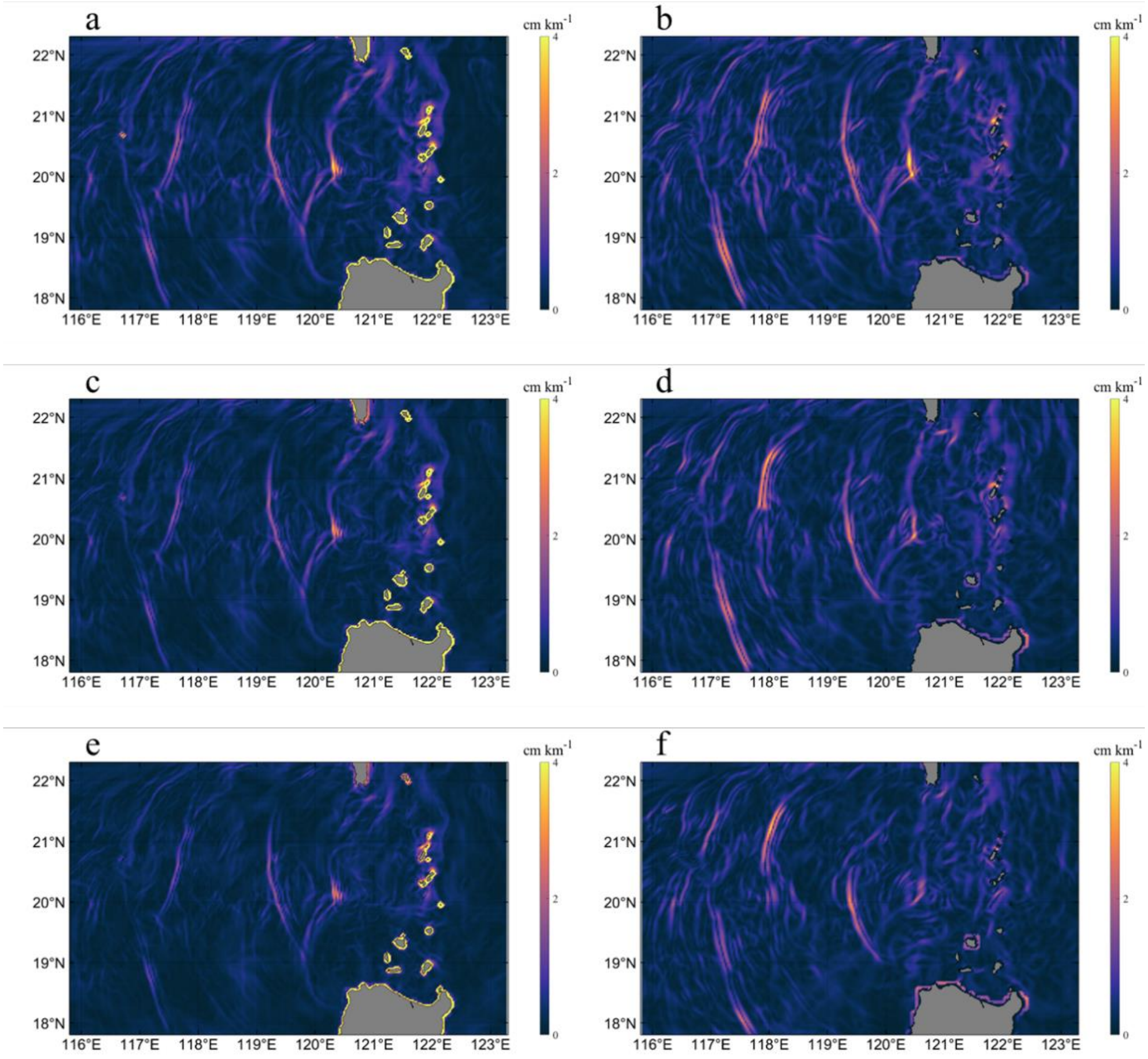

Figure 6 The same as Figure 5 except for the SSH gradients.

The energy spectrum distribution across different scales with implication of the energy cascades (particularly between the mesoscale and submesoscale) usually serves as a crucial physical criterion when evaluating the capability of an ocean model in representing submesoscale processes. Figure 7 shows the power spectra of SSH in the northern SCS fromVolador 1.0 for different forecast periods (0-24h, 25-48h, and 49-72h) averaged over the three months, as well as those from SWOT, REDOS V2.0 and GLORYS12 datasets. It should be noted that, the geographic coverage of SWOT data, derived from the along-track satellite measurements, does not fully coincide with those of the other datasets for the calculation of power spectrum, which may introduce additional discrepancies in the power spectrum for the comparison. As shown in Figure 7, the overall power spectra from Volador 1.0 for different forecast periods, REDOS V2.0 and SWOT, which all have higher resolution (~2-3km), are lager, flatter, and extends over a broader wavenumber range than that from GLORYS12 with a lowest resolution (~10km). In the regime of large scales (>100km), the power spectra from Volador 1.0 for different forecast periods closely align with that from REDOS V2.0, and exhibit a steeper slope compared to that from SWOT, following a $K^{-5}$ power law. This steeper slope pattern may be attributed to the difference of the geographic coverage between SWOT and the other two datasets. In the regimes of mesoscales (10km-100km) and submesoscales (<10km), the power spectra from both Volador 1.0 and REDOS V2.0 align with that from SWOT following a power law between $K^{-5}$ and $K^{-11/3}$, demonstrating their consistency with the classical QG turbulence theory of enstrophy cascading from submesoscale to mesoscale; in contrast, the spectrum from GLORYS12 shows a much deeper slope in the regime of mesoscales with weaker energy and lacks in the regime of submesoscales. When looking into the details, the spectra from REDOS V2.0 follows more closely to that from SWOT than those from Volador 1.0 that are relatively steeper with smaller magnitudes and decrease with the forecast time, especially in the region of mesoscales. This is likely because Volador 1.0 filtered out some of the high-frequency energy signals present in REDOS V2.0 during its inference process. Moreover, the spectrum from SWOT becomes much flatter with smaller scales in the regime of submesoscale, which may be due to the high-frequency, small-scale “noise" in SWOT. Overall, the close alignment of energy spectra from Volador 1.0 with those from SWOT and REDOS V2.0 across multiple scales, which are consistent with theoretical power laws, demonstrates Volador 1.0’s capability in correctly representing the energy distribution and cascades across different scales, especially the forward cascade from the sub- to mesoscale.

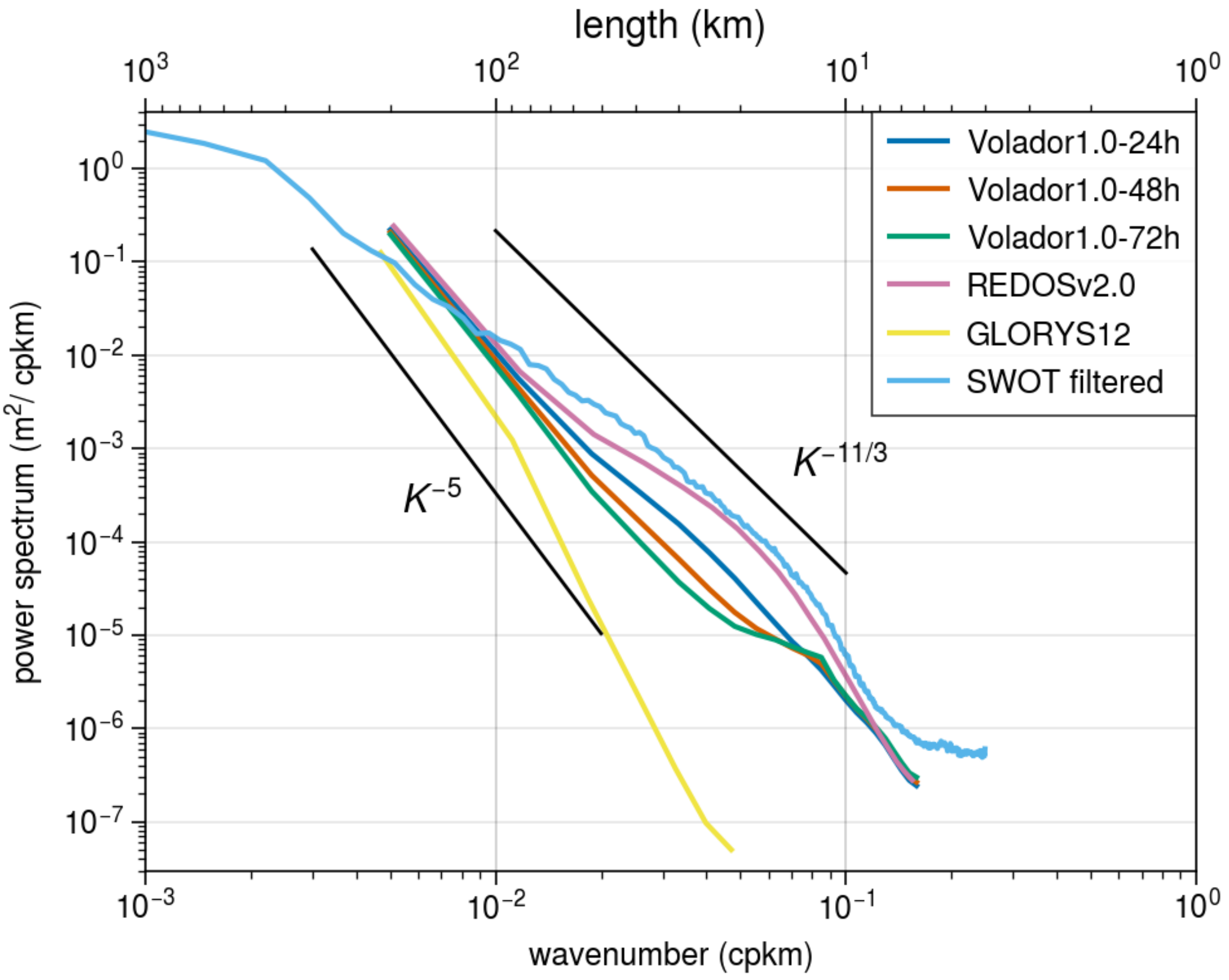

Figure 7 The power spectrum calculated based on SWOT-derived along-track SLA, REDOS V2.0, GLORYS, and Volador 1.0 with 0-24h, 25-48h and 49-72h forecasts averaged from June 1, 2024 to August 31, 2024.

**c. Benefits of air-sea full-coupling**

Momentum and heat exchanges across the air-sea interface are not only crucial for driving upper ocean evolution, but also profoundly regulate the vertical stratification of the upper ocean. To quantify the benefits that Volador 1.0 gains for its forecast skills from the air-sea full-coupling (i.e., “bidirectional-coupling”), two ablation experiments, in which the atmospheric forcings on the ocean surface are either fixed (i.e., “one-directional-coupling” through retaining atmospheric inputs but disabling the air-sea full-coupling module, denoted as Atmos_fixed) or not taken into account (i.e., without any exchange between the atmosphere and ocean through completely blocking the input of atmospheric feature tensors, denoted as Ocean_only), are carried out. It should be noted that the model with either of the ablation experimental configurations regarding to the atmospheric forcings is trained independently. To ensure the validity and fairness of the evaluation, the ablation experiments are conducted with the same daily rolling forecast scheme from June 1 to August 31, 2024 as the hindcast test.

Figures 8-9 show the time series of domain-mean vertical averaged RMSEs of temperature, salinity, zonal current, and meridional current as well as the domain-mean vertical RMSE profiles of temperature and salinity from the two ablation experiments and Volador 1.0 validated against REDOS V2.0. It can be found that the complete absence of thermodynamic and momentum forcings from the upper atmosphere with only relying on the ocean's internal autoregressive evolution (Ocean_only) results in much larger RMSEs compared to those with atmospheric forcings (Atmos_fixed and Volador 1.0), confirming the importance and necessity of including the atmospheric forcings in correctly simulating/forecasting the ocean state. Nevertheless, for the Atmos_fixed experiment with atmospheric forcings fixed (i.e., one-directional-coupling) during all the time of forecasting, the RMSEs are larger than those from Volador 1.0 with air-sea full-coupling after 24-hour forecasting, due to the absence of cross-modal bidirectional dynamic feedback which could have more impacts with the increase of forecast lead time. For the forecast period of 49-72h, Ocean_only and Atmos_fixed have about 22.3%–37.1% and 4.5%–7.5% increases in the domain-mean vertical averaged RMSEs of temperature, salinity, zonal current, and meridional current compared to Volador 1.0, respectively, as seen in Table 2. It is worth noting that the vertical RMSE profiles of temperature and salinity from Ocean_only exhibit a prominent peak in the subsurface layer corresponding to the thermocline with highly active physical gradients (Figure 9a), which are reduced significantly in both Atmos_fixed with air-sea one-directional-coupling and Volador 1.0 with air-sea bidirectional-coupling, confirming the importance of the atmospheric momentum input that control some of the key processes in the subsurface, i.e., wind-driven vertical turbulent mixing, without which would lead to severe misestimations of both the mixed layer and thermocline depths. And compared to the air-sea one-directional-coupling, the air-sea bidirectional-coupling gains more reduction of the RMSE peak in the subsurface, owning to its cross-modal bidirectional dynamic feedback between the atmosphere and ocean.

The above results indicate that Volador 1.0 benefits from the air-sea full-coupling framework that correctly represents the bidirectional air-sea interaction. Its underlying air-sea full-coupling layer extracts high-dimensional fused features rich in dynamic forcing and feeds them into a decoder with a "fast-slow dual-branch decoupling" structure. This enables Volador 1.0 to produce dynamic forcing fields that obey fluid physics laws, thereby greatly improves oceanic forecast skills.

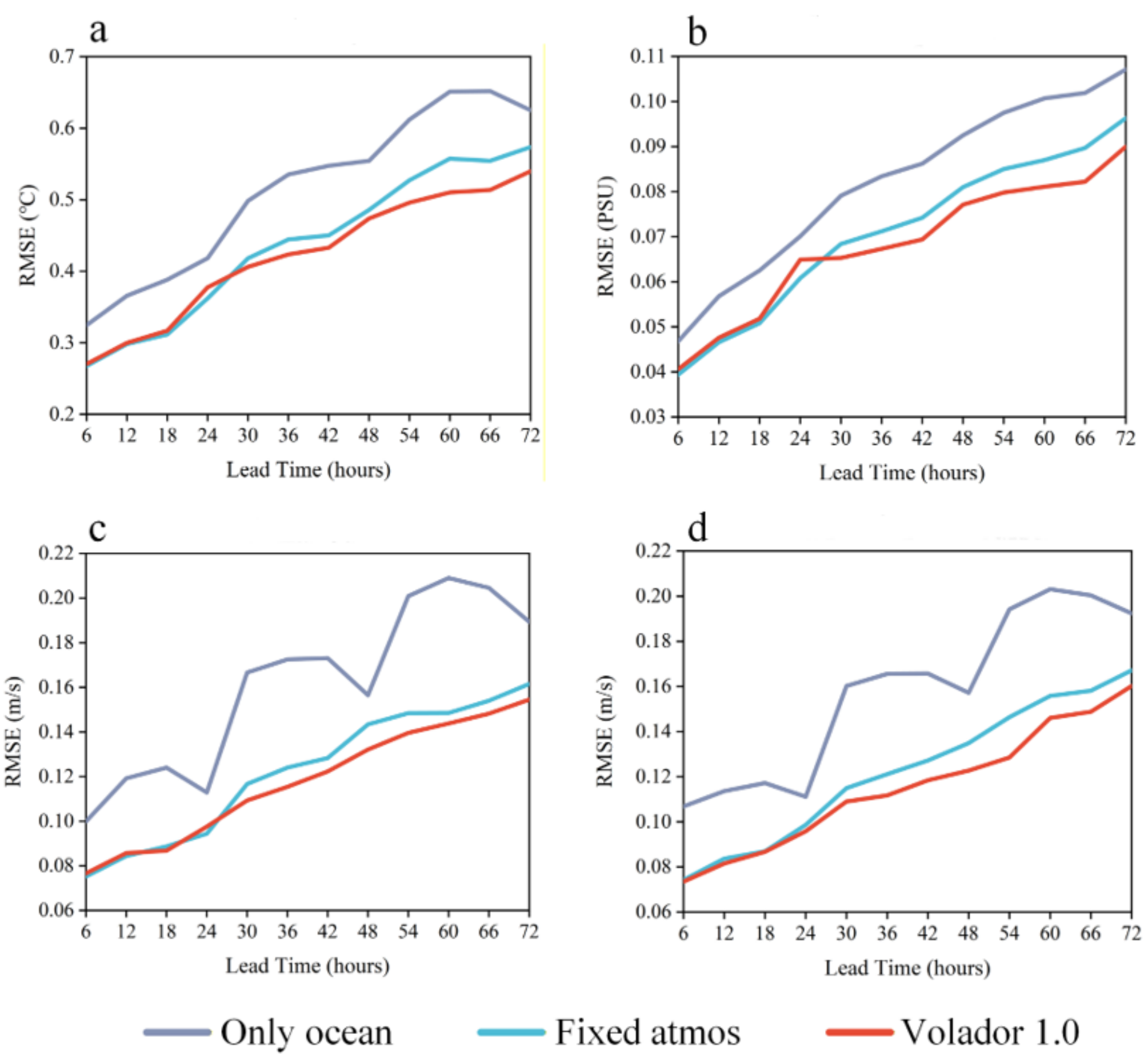

Figure 8 The time series of domain-mean vertical-averaged RMSEs of temperature (a), salinity (b), zonal current (c), and meridional current (d) from different ablation experiments and Volador 1.0 averaged from June 1, 2024 to August 31, 2024.

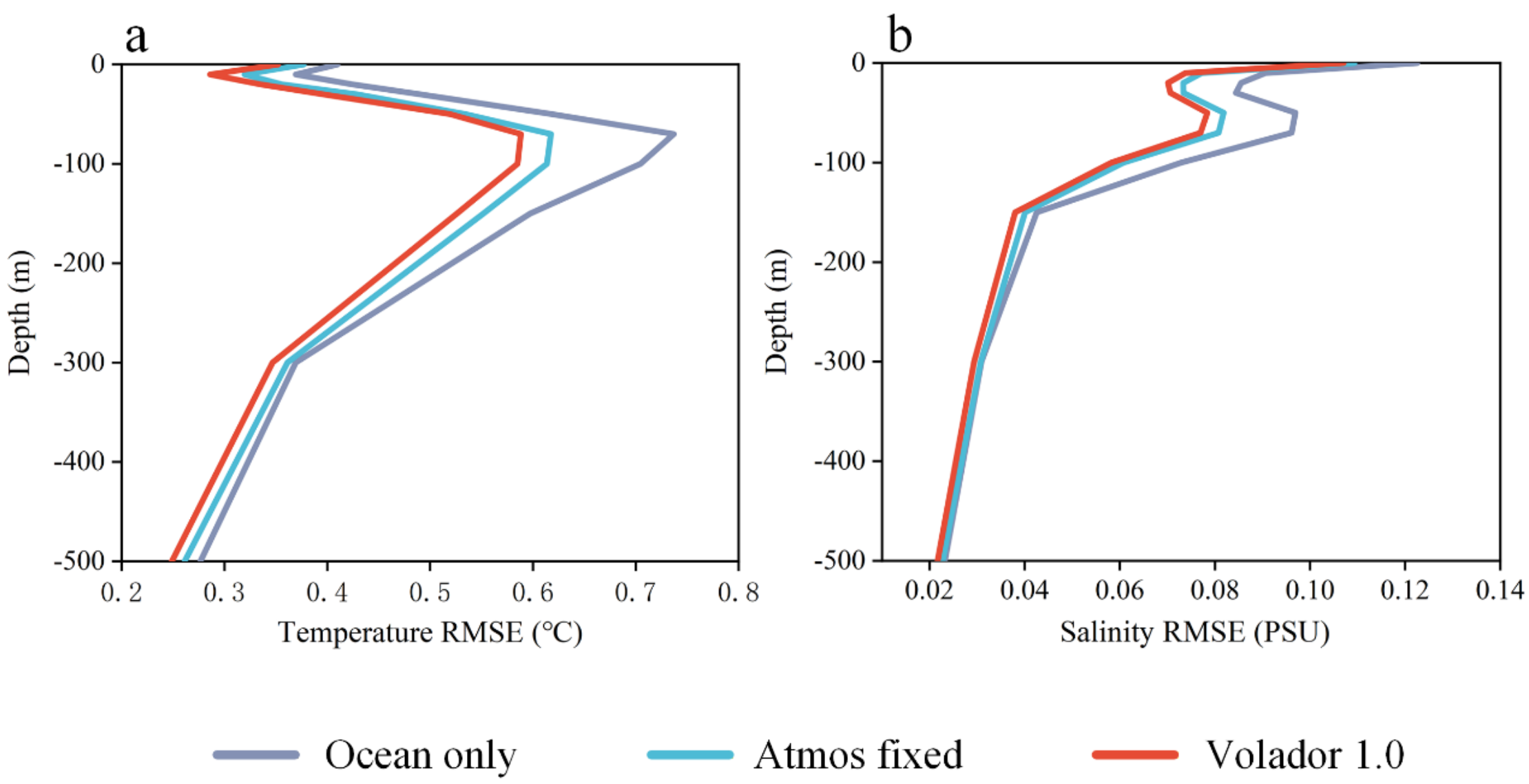

Figure 9 The domain-mean RMSEs of temperature (a) and salinity (b) profiles from different ablation experiments and Volador 1.0 averaged from June 1, 2024 to August 31, 2024.

Table 2 The domain-mean vertical-averaged RMSEs of temperature, salinity, u and v from 0-72h forecasts of Volador 1.0 and the ablation experiment averaged from Juan 1 2024 to August 31 2024. The values in parentheses indicate the percentage increase in the RMSE of the specific variable at 49-72h forecast period relative to Volador 1.0.

| Variable / Experiment | Temperature (℃) | Salinity (PSU) | U (m/s) | V (m/s) |
|---|---|---|---|---|
| Volador 1.0 | 0.422 | 0.068 | 0.118 | 0.115 |
| Atmos_fixed | 0.438 (7.4%) | 0.071 (7.5%) | 0.122 (4.5%) | 0.122 (7.5%) |
| Ocean_only | 0.514 (23.3%) | 0.082 (22.3%) | 0.161 (37.1%) | 0.157 (35.4%) |

### 4.2 Evaluation of Volador 1.0 on operational real-time forecasting

Being installed in an Intel GPU chip of A100, Volador 1.0 has just been implemented in the operational real-time forecasting, with input data sourced from the Real-time Forecast System for the SCS and the northwestern Pacific (RFSSW) developed by our research team. The atmospheric and oceanic models employed in RFSSW are the Weather Research and Forecasting (WRF) (Skamarock et al., 2008) and the ROMS, respectively, sharing the same domain in the northwestern Pacific that fully covers the domain of Volador 1.0. The horizontal resolution and vertical layers for WRF are 9km and 50, while those for ROMS are 1/60° ×1/60° and 50. The initial and lateral boundary conditions with 3-h interval for WRF are the hourly outputs from the Global Forecast System (GFS) with a horizontal resolution of 1/4° × 1/4° maintained by National Centers for Environmental Prediction (NCEP.). For ROMS, the outputs from the WRF provides the atmospheric forcing, and the multi-year monthly mean HYbrid Coordinate Ocean Model (HYCOM) reanalysis dataset (Cummings, 2005; Cummings and Smedstad, 2013) and tidal level/currents calculated based on the TPXO8 dataset are used as the lateral boundary conditions as well as the initial conditions of model spinup. The T/S profiles from Argo buoys and near real-time gridded satellite-derived SLA and SST are assimilated into ROMS when initializing the forecasting. The RFSSW initiates one forecast cycle per day at 00:00 UTC, providing the atmospheric and oceanic states for the upcoming 168 hours (7 days).

For a 72-h operational forecasting cycle of Volador 1.0 initialized at 00:00 UTC of a day, the input data are derived from the 6h, 12h, 18h, and 24h forecasting outputs of RFSSW initialized at 00:00 UTC of the previous day. It only needs 3 seconds to finish a 72-h forecasting running on the single GPU chip. Here, as a very preliminary evaluation, the T/S in the depth of 0-500m operationally forecasted by Volador 1.0 initialized at 00:00 UTC of each day from April 10 to April 24, 2026 are validated against the T/S profiles from Argo buoys (Figure 1). Figure 10 displays 15-day-mean RMSEs of T/S profiles averaged over the 0-72h forecast period from Volador 1.0 and RFSSW. It can be seen that the operational real-time forecast performance of Volador 1.0 is very encouraging, with T/S RMSEs of 0.925 °C and 0.165 PSU, respectively, compared to 1.084 °C and 0.181 PSU from RFSSW. Though further evaluation on more forecasting variables and longer forecasting period (say, half to one year) is absolutely needed, especially regarding to the aspect of performance stability, this preliminary result shows a great promise of the practical utility of Volador 1.0, which could contribute undoubtedly to the hazard-prevention and loss-reduction in the coastal regions of the SCS.

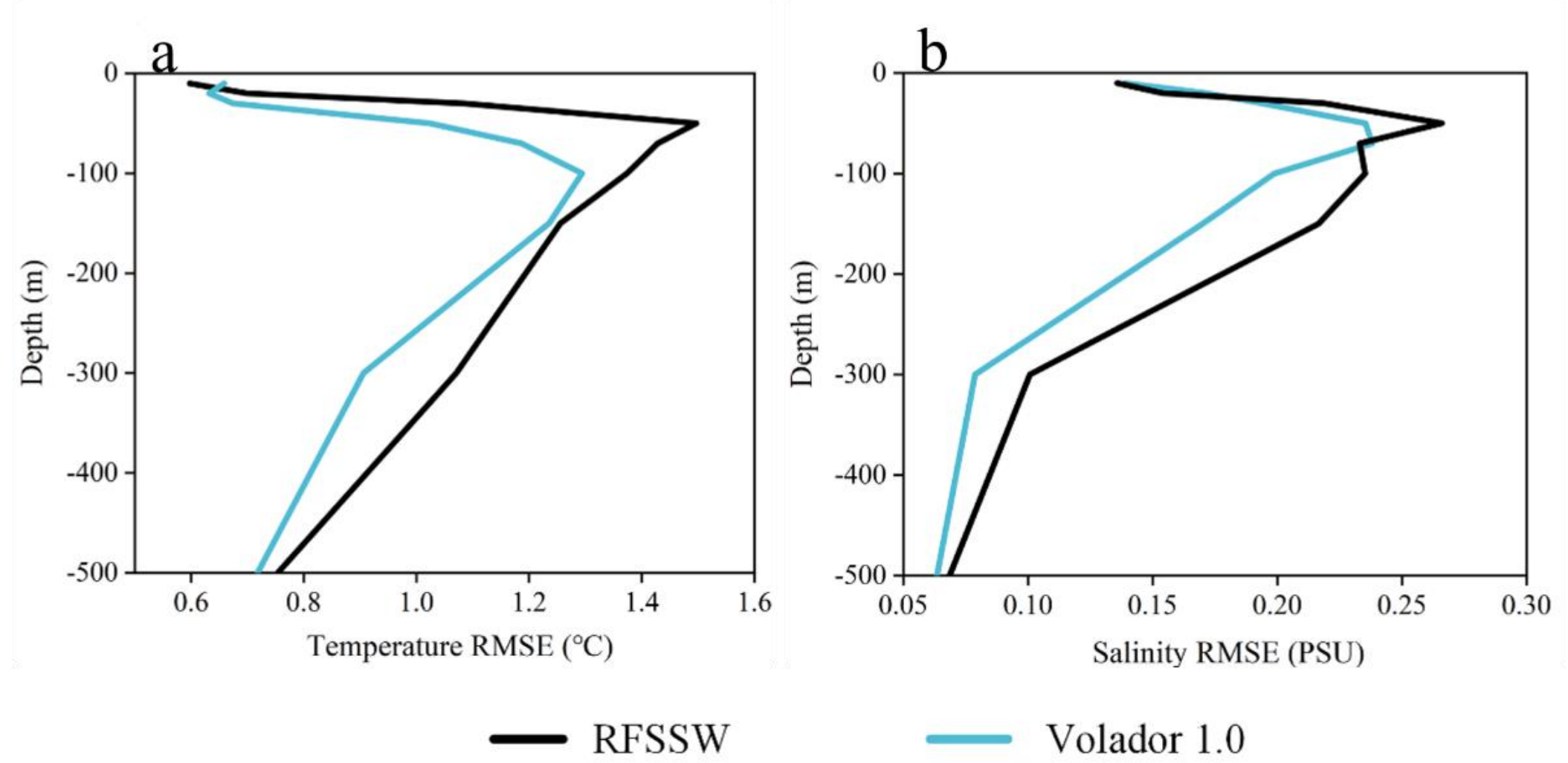

Figure 10 The domain-mean RMSEs of 0-72h forecasting temperature (a,) and salinity (b) profiles in the SCS (a, b) and the NWPO (c, d) from Volador 1.0 and RFSSW averaged from April 10 to April 25, 2026.

## 5 Summary

In this study, a data-driven air-sea full-coupling regional forecast model named "Volador 1.0" with submesoscale-permitting is developed for the SCS. Built on a Swin-Transformer framework integrated with a MoE system,Volador 1.0 divides complex atmosphere and ocean forecasting tasks into specialized expert sub-tasks via dynamic routing and activates them only when and where they are needed, which leverages its sparse activation characteristics to require only a minimal number of activation parameters during the inference stage, thereby achieving fine-grained forecasting of multi-scale, highly non-linear ocean dynamical processes and significantly reducing inference power consumption. Besides, the model introduces an air-sea full-coupling module which innovatively proposes a latent space interaction architecture based on Cross-Grid Bidirectional Cross-Attention, thereby establishing an efficient information exchange channel for dynamically forecasting atmospheric and oceanic states. As confirmed by the ablation experiments, Volador 1.0 obviously benefits from this air-sea full-coupling configuration which correctly simulates the momentum and heat flux transfer between atmosphere and ocean, leading to a good forecast skill. Moreover, considering that different oceanic variables exhibit significantly distinct physical evolution rates, a fast-slow dual-branch architecture is innovatively introduced to avoid gradient conflicts and feature over-smoothing problems when reconstructing these disparate variables simultaneously, which is based on physical quantity decoupling within the decoder stage by splitting the feature reconstruction process into two parallel computational pathways. Both the air-sea full-coupling configuration and the fast-slow dual-branch architecture of Volador 1.0 are based on the physics of the air-sea interactions and oceanic evolution, making Volador 1.0 an advanced data-driven model in physical interpolation. The input data used for training and testing Volador 1.0 are ERA5 dataset with horizontal resolution of 1/4°×1/4° for the atmospheric component and REDOS V2.0 with horizontal resolution of 1/36°×1/36° for the oceanic component, respectively; the very high resolution of the latter makes Volador 1.0 possess the capability of capturing/forecasting the submesoscale processes in the ocean.

As the first air-sea full-coupling model and the first oceanic-submesoscale-permitting model among all the state-of-the-art data-driven atmospheric or oceanic models in the world, Volador 1.0 demonstrates a very encouraging and promising performance in forecasting accuracy, submesoscale representation and efficiency of data-digging and inference as follows:1) having a very good forecast skill for ocean state with the RMSEs of 0-500m averaged temperature (salinity) being only 0.521℃, 0.558℃ and 0.614℃ (0.102PSU, 0.105PSU and 0.110PSU) and the MAEs of the SLA being 0.08m, 0.07m and 0.07m for 24h, 48h and 72h forecast periods, respectively, which are comparable to those from the reanalysis datasets REDOS V2.0 (with corresponding T/S RMSEs of 0.546℃ and 0.102PSU and SLA MAE of 0.08m) and GLORYS12 (with corresponding T/S RMSEs of 0.717℃ and 0.130PSU and SLA MAE of 0.08m), which had assimilated satellite-derived and in situ observations; 2) reasonably capturing/forecasting the submesoscale signals including internal waves and correctly representing the oceanic energy cascade from sub- (~a few kilometers) to mesoscale (10-100 km), which aligns with the classical turbulence theory; 3) using only 4.5-year data with an interval of 4-time/day for training but still outperforming the numerical model ROMS as demonstrating in operational real-time forecasting, as compared to at least 20-year daily data required by nearly all state-of-the-art data-driven atmospheric or oceanic models, and needing only a very short moment (~ 3 seconds) for its inference to make a 72-h real-time forecasting for the SCS on a single Intel GPU chip of A100, thereby not only implying the super ability of Volador 1.0 in extracting key features from large data but also undoubtedly reducing the power cost during both training and inference and thus providing a timely forecasting or early warning for hazard mitigation.

In conclusion, Volador 1.0 blazes a path for an accurate, fine and fast marine forecasting by synergizing innovative deep learning architectures with high-resolution training data, thereby achieving accurate and physically consistent forecasting of the physical oceanic elements and submesoscale processes in the dynamically complex SCS, which could help promote our capability of disaster prevention and mitigation in the SCS as well as in other coastal regions where these innovative techniques are applied. Nevertheless, it is worth noting that several limitations in the current study: 1) relatively few number of vertical layers in both the atmospheric and oceanic components, as well as relatively short data length for training and short forecast lead time; 2) no atmospheric and oceanic boundary update during forecasting which could lead to large biases in longer forecast lead times; 3) validation only for the oceanic forecast and leaving that for the atmospheric forecast to be done. These limitations will be our primary focus in the future to make an update from Volador 1.0 to Volador 2.0, which we do believe has a large space for the improvement of the model's performance compared to its current version.

## Data availability

For training and testing Volador 1.0, we downloaded the GLORYS12 dataset from https://data.marine.copernicus.eu/product/GLOBAL_MULTIYEAR_PHY_00L_030/services, ERA5 dataset from https://ods.climate.copernicus.eu/, Argo buoys dataset from https://www.argo-cndc.org/#/catalog/dataService and SWOT dataset from https://podaac.jpl.nasa.gov/SWOT. All these data are publicly available for research purposes.

## Code availability

The code base of Volador 1.0 was established on PyTorch, a Pythonbased library for deep learning. The details of Volador 1.0, including network architectures and modules, are available in the paper. We will release the inference model to the public at a GitHub repository: https://github.com/AnimaNet-team/Volador1.0. The inference model allows the researcher to explore Volador 1.0's ability.

## Acknowledgements

This work was jointly supported by the Strategic Priority Research Program of the Chinese Academy of Sciences (Grant XDA0500202).

## Author Contributions

S.Q. and T.S. designed the project. Y.Z. and J.W. developed the methodology. Y.Z., J.W., Y.Q. and Y.G. performed the model evaluation. YN.L. and SL.T. improved the model design. J. W. and YA. L. conducted the experiments. Y.Z. and J.W. wrote the original draft. S.Q. and Y.Q. reviewed and edited the mamscript.

## Competing Interests

The authors declare no competing interests.

## References

[1] AOUNI A E, GAUDEL Q, REGNIER C, et al. Glonet: Mercator's end-to-end neural forecasting system[PP/OL]. arXiv:2412.05454 (2024) [2026-05-06]. https://arxiv.org/abs/2412.05454

[2] BI K, XIE L, ZHANG H, et al. Accurate medium-range global weather forecasting with 3d neural networks[J]. Nature, 2023: 1-6.

[3] CHARNEY J G. Geostrophic turbulence. Journal of the Atmospheric Sciences, 1971, 28: 1087-1095.

[4] CHASSIGNET E P, HURLBURT H E, SMEDSTAD O M, et al. The HYCOM (HYbrid Coordinate Ocean Model) data assimilative system[J]. Journal of Marine Systems, 2007, 65: 60-83.

[5] CHEN G X, GAN J P, XIE Q, et al. Eddy heat and salt transports in the South China Sea and their seasonal modulations[J]. Journal of Geophysical Research: Oceans, 2012, 117: C05021. DOI: 10.1029/2011JC007724.

[6] CHEN G X, HOU Y J, CHU X Q. Mesoscale eddies in the South China Sea: mean properties, spatiotemporal variability, and impact on thermohaline structure[J]. Journal of Geophysical Research: Oceans, 2011, 116: C06018. DOI: 10.1029/2010JC006716.

[7] CUI Y, WU R, ZHANG X, et al. Forecasting the eddying ocean with a deep neural network[J]. Nature Communications, 2025, 16(1): 2268.

[8] CUMMINGS J A. Operational multivariate ocean data assimilation[J]. Quarterly Journal of the Royal Meteorological Society, 2005, 131(613): 3583-3604.

[9] CUMMINGS J A, SMEDSTAD O M. Variational data assimilation for the global ocean[M]//LEWIS J M, NAVON I M, ZUPANSKI M, et al. Data assimilation for atmospheric, oceanic and hydrologic applications(Vol II). Berlin, Heidelberg: Springer, 2013: 303-343.

[10] DAI A. Dai and Trenberth Global River Flow and Continental Discharge Dataset[DS]. Research Data Archive at the National Center for Atmospheric Research, Computational and Information Systems Laboratory (2017) [2020-12-25]. https://doi.org/10.5065/D6V69H1T

[11] D'ASARO E, LEE C, RAINVILLE L, et al. Enhanced turbulence and energy dissipation at ocean fronts[J]. Science, 2011, 332(6027): 318-322. DOI: 10.1126/science.1201515.

[12] EGBERT G D, EROFEEVA S Y. Efficient inverse modeling of barotropic ocean tides[J]. Journal of Atmospheric and Oceanic Technology, 2002, 19(2): 183-204.

[13] FANG G, FANG W, FANG Y, et al. A survey of studies on the South China Sea upper ocean circulation[J]. Acta Oceanographica Taiwanica, 1998, 37: 1-16.

[14] GUO P, FANG W, LIU C, et al. Seasonal characteristics of internal tides on the continental shelf in the northern South China Sea[J]. Journal of Geophysical Research, 2012, 117: C04023. DOI: 10.1029/2011JC007215.

[15] HU J, KAWAMURA H, HONG H, et al. A review on the currents in the South China Sea: seasonal circulation, South China Sea warm current and Kuroshio intrusion[J]. Journal of Oceanography, 2000, 56: 607-624.

[16] LELLouche J M, GREINER E, BOURDALLÉ-BADIE R, et al. The Copernicus global 1/12° oceanic and sea ice GLORYS12 reanalysis[J]. Frontiers in Earth Science, 2021, 9: 698876.

[17] LI J X, ZHANG R, JIN B G. Eddy characteristics in the northern South China Sea inferred from Lagrangian drifter data[J]. Ocean Science, 2011, 7: 661-669.

[18] LIU Q, KANEKO A, JILAN S. Recent progress in studies of the South China Sea circulation[J]. Journal of Oceanography, 2008, 64: 753-762.

[19] LELLOUCHE J M, GREINER E, BOURDALLE-BADIE R, et al. The Copernicus Global 1/12° Oceanic and Sea Ice GLORYS12 Reanalysis[J]. Frontiers in Earth Science, 2021, 9: 698876. DOI: 10.3389/feart.2021.698876.

[20] MA B B, LIEN R C, KO D S. The variability of internal tides in the northern South China Sea[J]. Journal of Oceanography, 2013, 69: 619-630. DOI: 10.1007/s10872-013-0198-0.

[21] WEN M Q, QING C X, FANG Y Y, et al. A three-dimensional temperature and salinity reconstruction system in the South China Sea[J]. Journal of Tropical Oceanography, 2013. DOI: 10.3969/j.issn.1009-5470.2013.06.001.

[22] MISHONOV A V, BOYER T P, BARANOVA O K, et al. World Ocean Database 2023[DS]. NOAA Atlas NESDIS 97. Silver Spring, MD: NOAA National Centers for Environmental Information, 2024: 206. DOI: 10.25923/z885-h264.

[23] MIYAZAWA Y, VARLAMOV S M, MIYAMA T, et al. Assimilation of high-resolution sea surface temperature data into an operational nowcast/forecast system around Japan using a multi-scale three-dimensional variational scheme[J]. Ocean Dynamics, 2017, 67: 713-728.

[24] LAM R, SANCHEZ-GONZALEZ A, WILLSON M, et al. Learning skillful

medium-range global weather forecasting[J]. Science, 2023: eadi2336.
[25] REAGAN J R, BOYER T P, GARCÍA H E, et al. World Ocean Atlas 2023[DS]. NOAA National Centers for Environmental Information, 2024. Dataset: NCEI Accession 0270533.
[26] SKAMAROCK W C, KLEMP J B. A time-split nonhydrostatic atmospheric model for weather research and forecasting applications[J]. Journal of Computational Physics, 2008, 227: 3465-3485.
[27] SMEDSTAD O M, HURLBURT H E, METZGER E J, et al. An operational real-time eddy-resolving 1/16° global ocean nowcast/forecast system[J]. Journal of Marine Systems, 2003, 40: 341-361.
[28] QU T D. Upper-layer circulation in the South China Sea[J]. Journal of Physical Oceanography, 2000, 30: 1450-1460.
[29] SONG T, HAN N, ZHU Y, et al. Application of deep learning technique to the sea surface height prediction in the South China Sea[J]. Acta Oceanologica Sinica, 2021, 40(7): 1-9. DOI: 10.1007/s13131-021-1735-0.
[30] SONG Y T. Estimation of interbasin transport using ocean bottom pressure: theory and model for Asian marginal seas[J]. Journal of Geophysical Research, 2006, 111: C11S19. DOI: 10.1029/2005JC003189.
[31] WANG D, LIU Q, HUANG R X, et al. Interannual variability of the South China Sea throughflow inferred from wind data and an ocean data assimilation product[J]. Geophysical Research Letters, 2006, 33: L14605. https://doi.org/10.1029/2006GL028103
[32] WANG D X, LIU Q Y, XIE Q, et al. Progress of regional oceanography study associated with western boundary current in the South China Sea[J]. Chinese Science Bulletin, 2013, 58: 1205–1215.
[33] WANG X, WANG R, HU N, et al. Xihe: A data-driven model for global ocean eddy-resolving forecasting[PP]. Beijing: arXiv, 2024. https://arxiv.org/abs/2402.02995
[34] XIONG W, XIANG Y, WU H, et al. Ai-goms: Large ai-driven global ocean modeling system[PP]. Beijing: arXiv, 2023. https://arxiv.org/abs/2308.03152
[35] YU Z, SHEN S, McCREARY J P, et al. South China Sea throughflow as evidenced by satellite images and numerical experiments[J]. Geophysical Research Letters, 2007, 34: L01601. https://doi.org/10.1029/2006GL028103
[36] ZHANG C L, XU J P, LIU Z H, et al. User manual of three-dimensional grid

dataset (GDCSM_Argo)[DS]. Hangzhou: China Argo Real-time Data Center, 2018. http://www.argo.org.cn/index.php?m=content&c=index&f=lists&catid=32

[37] ZHAO Z. Internal tide radiation from the Luzon Strait[J]. Journal of Geophysical Research: Oceans, 2014, 119: 5434–5448. https://doi.org/10.1002/2014JC010014